\documentclass[prb,twocolumn]{revtex4-1}

\usepackage{amsmath}  
\usepackage{amsfonts} 
\usepackage{graphicx} 
\usepackage{color}

\newcommand {\tauc} {\tau_\mathrm{c}}

\begin{document}


\title{
Connecting field and intensity correlations: the Siegert relation and how to test it
}

\author{Dilleys Ferreira}
\affiliation{CAPES Foundation, Ministry of Education of Brazil, Caixa Postal 250, Brasília  DF 70040-020, Brazil}
\affiliation{Universit\'e C\^ote d'Azur, CNRS, INPHYNI, France}

\author{Romain Bachelard}
\affiliation{Departamento de F\'{\i}sica, Universidade Federal de S\~{a}o Carlos, Rodovia Washington Lu\'{\i}s, km 235 - SP-310, 13565-905 S\~{a}o Carlos, SP, Brazil}
\affiliation{Universit\'e C\^ote d'Azur, CNRS, INPHYNI, France}

\author{William Guerin}
\author{Robin Kaiser}
\author{Mathilde Fouch\'e}
\email{mathilde.fouche@inphyni.cnrs.fr}
\affiliation{Universit\'e C\^ote d'Azur, CNRS, INPHYNI, France}

\date{\today}

\begin{abstract}
The Siegert relation relates the electric field and intensity correlations of light, under given assumptions. After a brief history of intensity correlations, we give a derivation of the relation. Then we present an experiment, which can be easily adapted for an undergraduate setup, and that allows measuring both field and intensity correlations at the same time, thus providing a direct test of the Siegert relation. As a conclusion, we discuss typical situations where the relation fails.
\end{abstract}

\maketitle

\section{Introduction}

Coherence properties of light is one of the pillars in modern optics lectures, and this subject is usually addressed during the undergraduate level. Coherence gives access to different kinds of information, on the light itself with its coherence time or the size of the source through the measurement of its spatial coherence\,\cite{Mandel1965}, on the emission processes such as the thermal or chaotic one in a light bulb or the stimulated emission in a laser, or even on the light-matter interaction when one studies for example the light scattered by atoms or molecules\,\cite{Maret_1987,Pine_1988,Eloy_2018}.

Those coherence properties can be characterized in different ways. Among the most used quantities, one can first cite the temporal field correlation function $g^{(1)}(\tau)$, also called first-order correlation function, defined as:
\begin{equation}\label{eq:g1}
    g^{(1)}(\tau) = \frac{\langle E^\star(t) E(t + \tau) \rangle}{\langle I(t) \rangle},
\end{equation}
with $\langle . \rangle$ corresponding to the averaging over the time $t$, and $I(t) = E^\star(t)E (t)$ the intensity associated to the field $E(t)$. Throughout this work, we consider stationary processes, so the dependence on the time $t$ is considered implicit, whereas the one over $\tau$ is left explicit. The function $g^{(1)}(\tau)$ is equal to 1 for $\tau = 0$ by definition and goes to 0 for large $\tau$ when the electric fields are completely uncorrelated. The typical decay time corresponds to the coherence time $\tau_\mathrm{c}$ of the electric field. It can be computed by measuring the contrast of the interference in a Michelson interferometer, adjusting the delay $\tau$ through the length imbalance between the two arms. As discussed later, one can also record the beat note between the light under study and another electric field with well-known coherence properties or with a coherence time much larger than the one we want to measure\,\cite{Razdan_2002}. The second well known quantity is the optical spectrum $S(\omega)$, or electric field spectrum, related to the degree of first-order coherence through the Wiener-Khintchine theorem \cite{Wiener1930,Khintchine1934} for stationary random processes:
\begin{equation}
    S(\omega) = \int g^{(1)}(\tau) e^{i\omega\tau} d\tau.
\end{equation}
It relates the temporal statistical properties of the fluctuated field to the results obtained with spectroscopic experiments. Finally, one can also measure the frequency noise power spectral density (PSD), using a frequency to intensity convertor such as a Fabry-Perot cavity\,\cite{Bava2002,Kashanian_2016}. This quantity is less exploited than the spectrum. Nevertheless, while the optical spectrum allows us to quickly compare different types of sources, the frequency noise PSD actually gives a more complete knowledge of the spectral properties.

All the coherence quantities cited in the previous paragraph (field correlation function, optical spectrum, frequency noise) are linked to the first-order correlation function of the electromagnetic field. However, since the work of Glauber\,\cite{Glauber:1963b}, it is  well known that if one aims at fully characterizing the coherence properties of a light source, one needs to measure the correlation functions $g^{(n)}$ at all orders $n$. Going one step further, one gets the second-order temporal correlation function, also called temporal intensity correlation function:
\begin{equation}\label{eq:g2}
    g^{(2)}(\tau) = \frac{\langle I(t)I(t + \tau) \rangle}{\langle I(t) \rangle^2}.
\end{equation}
When the electric field is treated as a quantum operator, as it is the case, for example, for the fluorescence spectrum of two-level atoms~\cite{Loudon:book}, $\langle . \rangle$ refers to expectation value, and the intensity correlation function should follow normal ordering. We briefly return on this point in Sec.\,\ref{sec:Siegert_derivation}.

For chaotic light, or more generally for fields that present properties of Gaussian processes, it happens that the correlation functions $g^{(n)}(\tau)$ are simply linked to each other\,\cite{VanKampen}. In particular, for spatially-coherent polarized chaotic light, the second-order correlation function $g^{(2)}(\tau)$ is related to the modulus of the temporal first-order correlation function $g^{(1)}(\tau)$ as follows:
\begin{equation}\label{eq:Siegert_simple}
    g^{(2)}(\tau) = 1+|g^{(1)}(\tau)|^2.
\end{equation}
This equation is commonly called the Siegert relation~\cite{Siegert:1943}. A straightforward consequence is that there is an excess of intensity correlation at zero delay, $g^{(2)}(0)>g^{(2)}(\tau\rightarrow \infty) = 1$. This is known as ``photon bunching'' or the ``Hanbury~Brown and Twiss'' effect for historical reasons that will be detailed in Sec.~\ref{subsec:history}. The Siegert relation is particularly useful when combined with the Wiener-Khintchine theorem, as it provides a direct link between the spectrum linewidth of the light and the correlation time of its intensity fluctuations. This connection has been used in different fields, from astronomy\,\cite{HBT:1956b,Guerin:2017} to dynamic light scattering\,\cite{Maret_1987,Pine_1988,Eloy_2018}.

Finally, it is well known in quantum optics that the Siegert relation is not necessarily valid. For example, laser light has a flat uncorrelated intensity correlation function\,\cite{Loudon:book}. Its intensity fluctuations are only due to the shot noise. In this kind of situation, measuring $g^{(2)}(\tau)$ opens the way to the study quantum effects and allows us to discriminate between different types of sources and to classify them.

Undergraduate experiments are usually designed to measure either quantities related to the first-order correlation function, such as the temporal correlation function or the spectrum, or the temporal intensity correlation function\,\cite{Basano_1982,Goldburg_1999,Kuusela_2017}. The purpose of this paper is to show how to slightly modify those experiments to measure the two quantities $g^{(1)}(\tau)$ and $g^{(2)}(\tau)$ at the time, and to test the validity of the Siegert relation.
After a quick physical picture and a brief history, we first derive the Siegert relation in Sec.\,\ref{sec:Siegert_derivation} in the specific case of light emitted by a large number of uncorrelated emitters. Then, in Sec.\,\ref{sec:setup} we describe the experimental setup, which have been implemented on our cold atom experiment. We present in Sec.\,\ref{sec:Tests} some results obtained for light scattered by the atoms from the single to the multiple scattering regime, with both low and strong driving fields. Finally, in the conclusion, we quickly give some well-known examples for which the Siegert relation is not valid.

\subsection{Intensity correlations explained in classical terms\label{subsec:classexplain}}

A derivation of the Siegert relation will be detailed in Sec.~\ref{sec:Siegert_derivation}. The purpose of this subsection is rather to give a simple physical picture to understand this relation in the simplest case of a spatially coherent chaotic source.

Let us consider a radiation with a finite optical spectrum and a linewidth $\Delta\omega$. The fact that the source is chaotic means that there is no phase relationship between the different spectral components. This is the case, for instance, if light comes from many independent emitters with different velocities, resulting in thermal light emission. In this configuration, the spectral phase $\phi(\omega)$ can be considered as completely random.

Let us now consider two frequency components from the optical spectrum. They induce a beat note at a frequency given by the difference of their optical frequencies. Since $\phi(\omega)$ is a random variable, all the possible beat notes coming from all possible pairs sum up with random phases. This is what generates intensity fluctuations. Note that this is a fully classical noise, due to the wave nature of the field and to its non-monochromaticity. This noise adds up to the photon noise (shot noise), which has a quantum origin.

If the linewidth of the spectrum is infinite, one would get white noise and thus an intensity correlation function equal to unity, whatever the delay $\tau$. On the other hand,
a finite linewidth means that there is no beating at a frequency much larger than $\Delta\omega$. This cut-off in the power spectrum of the noise corresponds to a finite correlation time $\tauc \sim 1/\Delta\omega$, and thus to correlated intensity fluctuations on this typical time scale. In other words, while $g^{(2)}(\tau\rightarrow \infty) = 1$ when intensity fluctuations are completely uncorrelated, one gets an increase of the correlation function at delays shorter than the correlation time. In the quantum realm this is referred as photon bunching.

In the spatial domain, a similar description can be given\,\cite{Hariharan_1980}. The interference of light coming from different points of the source gives rise to a speckle pattern. The decay length of the corresponding spatial intensity correlation function corresponds to the size of the speckle grain, which is proportional to $\lambda / \theta$, with $\lambda$ the central light wavelength and $\theta$ the angular size of the source, thus proportional to the inverse of the source size\,\cite{Goodman:book}.

\subsection{A brief history on intensity correlations}\label{subsec:history}

The history of the Siegert relation is intimately linked to the controversy on equal-time intensity autocorrelations, also known as the Hanbury~Brown and Twiss (HBT) effect. The story starts during World War II, when radar technology drove a lot of research in the field of radio waves with, later, much repercussion on radio astronomy and optical sciences. The relation between electric field correlations and intensity correlations has been proposed in that context by A. J. F. Siegert~\cite{Siegert:1943} in a report which is now unclassified. It was later named ``Siegert relation'', mainly in the field of mesoscopic physics\,\cite{Teich_1988,akkermans_montambaux_2007}.

\begin{figure}[ht]
	\centering
	\includegraphics[width=\columnwidth]{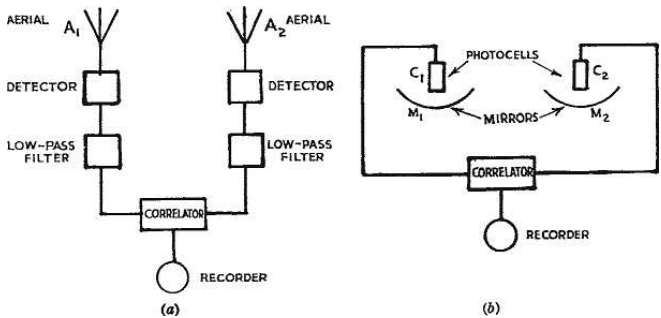}
	\caption{Simplified outline of an intensity interferometer for radio waves (a) and optical frequencies (b) taken from Ref.\,\cite{HBT:1956a}}
	\label{fig:HBT_1956}
\end{figure}

The next important step has been done by Hanbury~Brown and Twiss in the field of radio astronomy. In 1952, they proposed and demonstrated a novel type of radio interferometer, in which the intensities collected by two different telescopes pointing on the same star were correlated in time, as shown in Fig.\,\ref{fig:HBT_1956}, without recording the electromagnetic field phase information. Varying the distance between the two telescopes allowed them to record the spatial intensity correlation function from which the first measurements of the angular size of astronomical radio sources were extracted~\cite{HBT:1952,HBT:1954}. In their 1954 paper they wrote: `\textit{It is further shown that the correlator output, when suitably normalized, is equal to the square of the correlation coefficient measured by the Michelson interferometer}'. This statement links the intensity correlation function with the field correlation function, corresponding to the Siegert relation. Indeed the original technical report by Siegert had remained largely unnoticed and the relation was independently rediscovered at that time.

It was then natural, for radio-astronomers, to extend this new concept to visible light. However, they faced a strong opposition from several physicists who liked to think about light in terms of photons~\cite{HB:Boffin}. Indeed, a temporal (or spatial) intensity correlation measurement relies on the detection of at least two photons. The classical description given in the previous section only relies on interference. What was puzzling at that time is that physicists were convinced, following Dirac~\cite{Dirac:1930}, that `\textit{interference between two different photons can never occur}'.

HBT finally tested successfully their idea with a laboratory demonstration~\cite{HBT:1956a}, and a few months later on the light from the sky~\cite{HBT:1956b}. Their results were first disputed as other groups failed to reproduce the lab experiment, and as it was claimed that such results, if true, would call for a major revision of quantum mechanics~\cite{Brannen:1956}. Nevertheless, it was later shown that
the other experiments were simply not sensitive enough~\cite{HBT:1956c}. These first experiments were performed in the continuous regime, in which case there is no need to use the notion of photons and the classical explanation presented in Sec.~\ref{subsec:classexplain} is perfectly appropriate. In the photon counting regime, one can still assume that the quantization only occurs at the detection of an underlying continuous quantity, in which case the instantaneous value of the intensity $I(t)$ gives the probability of detecting photons, and the classical picture is still valid.

However, if one insists on describing light in terms of photons, another physical description is needed. A first argument was given by Purcell in 1956\,\cite{Purcell:1956}: the bunching of photons is a consequence of the Bose-Einstein statistics to which they obey. This interpretation was further developed by HBT~\cite{HBT:1957a} and Kahn~\cite{Kahn:1958}, and verified in an experiment done in the photon-counting regime~\cite{HBT:1957}. Finally, another interpretation in terms of two-photon interference was given by Fano a few years later~\cite{Fano:1961}.

The HBT experiment and its understanding in the framework of quantum mechanics can be considered as the birth of modern quantum optics (before the laser was invented!). In particular, and as acknowledged by Glauber in his Nobel speech~\cite{Glauber:2006}, it triggered the development of the quantum theory of optical coherence~\cite{Glauber:1963a,Glauber:1963b,Glauber:1963c}, based in particular on correlation functions. In this context, the Siegert relation, which provides a relation between first and second-order field correlations, is a particularly important tool to probe the quantum nature of light.

\section{Derivation of the Siegert relation}\label{sec:Siegert_derivation}

The Siegert relation is valid if one considers a complex random Gaussian field. A common situation in which such a field is encountered is the case of a large number of uncorrelated emitters. We hereafter derive the Siegert relation in this configuration.

\subsection{Electric field correlations}
\label{subsec:g1}

The electric field radiated by a collection of $N$ independent emitters (i.e., a chaotic light source) can be written as follows~\cite{Loudon:book,Chu2012}:
\begin{eqnarray}
    E(t) &=& \sum_{j=1}^N E_j(t),\\
    &=& E_0 e^{i \omega_0 t}\sum_{j=1}^N e^{i\phi_j(t)},
\end{eqnarray}
with $\omega_0$ the optical frequency, $E_0$ the positive amplitude of the field emitted by the different emitters and $\phi_j$ the phase due to emitter $j$. Out of simplicity, we assign the same fixed amplitude for each emitter, nevertheless this assumption is not crucial to derive the Siegert relation. The numerator of the first-order correlation $g^{(1)}$ introduced in Eq.\,(\ref{eq:g1}) reads:
\begin{eqnarray}
    \langle E^\star(t)E(t + \tau) \rangle &=& \sum_{j = 1}^N \sum_{k = 1}^N \langle E_j^\star(t)E_k(t + \tau) \rangle,\\
    &=& E_0^2 e^{i \omega_0\tau} \sum_{j = 1}^N \sum_{k = 1}^N \langle e^{-i[\phi_j(t)-\phi_k(t+\tau)]} \rangle,
    \\ &=& E_0^2 e^{i \omega_0\tau} \sum_{j = 1}^N \langle e^{-i[\phi_j(t)-\phi_j(t+\tau)]} \rangle.   \label{eq:g1_independent}
\end{eqnarray}
The crossed terms $j\neq k$ cancel since we consider independent emitters and thus uncorrelated phases (that is, $\langle e^{-i[\phi_j(t)-\phi_k(t+\tau)]} \rangle =0$). This can be due, for example, to a spectral broadening mechanism, such as the Doppler or collision effects. Using the same argument for the denominator, the first-order correlation function is thus given by:
\begin{eqnarray}
    g^{(1)}(\tau) &=& \frac{E_0^2 e^{i \omega_0\tau}\sum_{j = 1}^N \langle e^{-i[\phi_j(t)-\phi_j(t+\tau)]} \rangle}{NE_0^2},
    \\     &=& \frac{e^{i \omega_0\tau}}{N}\sum_{j = 1}^N \langle e^{-i[\phi_j(t)-\phi_j(t+\tau)]} \rangle.
\end{eqnarray}
As mentioned in the introduction, at zero delay, we get $g^{(1)}(\tau) = 1$, by definition. As $\tau$ is increased, $g^{(1)}(\tau)$ decreases to 0 and the decay time typically corresponds to the coherence time $\tau_\mathrm{c}$ of the electric field. According to the Wiener-Khintchine theorem, the temporal first-order correlation function also corresponds to the Fourier transform of the spectrum of the emitted light. This provides a direct connection between the linewidth of the spectrum and the inverse of the coherence time: $\Delta\omega\sim 1/\tauc$.

\subsection{Intensity correlation}
\label{subsec:g21}

The temporal intensity correlations can be expanded as follows:
\begin{widetext}
\begin{eqnarray}
    \langle I(t)I(t + \tau) \rangle
    = \sum_{j,k,l,m = 1}^N \langle E_j^\star(t)E_k(t) E_l^\star(t + \tau)E_m(t + \tau) \rangle = \sum_{j,k,l,m = 1}^N E_0^4 \langle  e^{-i[\phi_j(t)-\phi_k(t)+\phi_l(t+\tau)-\phi_m(t+\tau)]} \rangle.\nonumber
\end{eqnarray}
In the above equation, using the same argument as for the field correlations, only the terms which obey one of the following pairwise equalities do not cancel over averaging:
\begin{itemize}
    \item If $j=k=l=m$, one obtains $\sum_j E_0^4=N E_0^4$,
    \item If $j=k$ and $l=m$, but $j\neq l$, one gets $\sum_{j,l\neq j}E_0^4= N(N-1) E_0^4$,
    \item If $j=m$ and $k=l$, but $j\neq k$, one gets $\sum_{j,k\neq j}E_0^2 \langle e^{-i[\phi_j(t)-\phi_j(t+\tau)]} \rangle E_0^2 \langle e^{i[\phi_k(t)-\phi_k(t+\tau)]} \rangle  = N(N-1) E_0^4 |g^{(1)}(\tau)|^2$.
\end{itemize}
\end{widetext}
As for the denominator of $g^{(2)}$, it is simply equal to $N^2 E_0^4$, which leads to the following expression for the second-order correlation:
\begin{equation}\label{eq:g2N}
    g^{(2)}(\tau) = \frac{1}{N} + \left(1-\frac{1}{N}\right)\left[1+\big|g^{(1)}(\tau)\big|^2\right].
\end{equation}

\subsection{Siegert relation}
\label{subsec:Siegert_relation}

If one now assumes that the number of emitters is large ($N\to\infty$), which is needed to assume that the total electric field obeys  Gaussian statistics, one obtains the Siegert relation:
\begin{equation}
    g^{(2)}(\tau) = 1+\big|g^{(1)}(\tau)\big|^2. \label{eq:Siegert_simple_bis}
\end{equation}
A factor $\beta \leq 1$ is sometimes added to take into account the reduction of the contrast as one averages the field or intensity over uncorrelated speckles or modes~\cite{Lemieux1999}:
\begin{equation}
    g^{(2)}(\tau) = 1+\beta\big|g^{(1)}(\tau)\big|^2.\label{eq:Siegert_beta}
\end{equation}
Experimentally, we will collect one spatial mode using a monomode fiber and select one polarization with a polarizer. In this configuration, $\beta = 1$ and the contrast is maximum. Finally, the Siegert relation can be also given in the Fourier space:
\begin{equation}
    \tilde{g}^{(2)}(\omega) = \delta(0)+\tilde g^{(1)}(\omega) * \tilde g^{(1)\star}(\omega),
\end{equation}
where $\tilde{g}$ refers to the Fourier transform, $*$ to the convolution, and $\tilde g^{(1)\star}$ to the conjugate of $\tilde g^{(1)}$.

Note that higher-order correlations $g^{(n)}$, corresponding to $n$-time intensity correlations ($n\geq 2$), have been already studied in the sixties and the seventies, in the context of laser light statistics~\cite{Chopra1973,Cantrell1973,Corti1974,Corti1976}, stellar interferometry~\cite{Mandel1965} and speckle noise~\cite{Labeyrie1970, Lohmann1983}. Indeed, assuming a Gaussian complex electric field implies that a relation between all the orders of correlations exist. Nevertheless, a rigorous generalization of the Siegert relation at higher orders was introduced only in the nineties ~\cite{Lemieux1999}.

Finally, a quantum treatment of the correlation functions $g^{(1)}$ and $g^{(2)}$ requires the use of electric field operator $\hat{E}$, carefully accounting for their ordering (also known as ``normal ordering''). Such an approach is required, for example, to describe properly the temporal correlations of the field radiated by an atom. Nevertheless, the above demonstration remains overall valid as soon as the product of the operators can be factorized as $\langle\hat{E}^\dagger_j(t)\hat{E}_k(t+\tau)\rangle = \langle\hat{E}^\dagger_j(t)\rangle \langle\hat{E}_k(t+\tau)\rangle$, which is true in particular for a large number of non-interacting atoms.
One can refer to Ref.~\cite{Loudon:book} for further details on the quantum treatment of optical coherences, and its comparison to the classical case.

\section{Experimental setup to test the validity of the Siegert relation}\label{sec:setup}

\subsection{Experimental setup}\label{subsec:setup}

\begin{figure}[ht]
	\centering
	\includegraphics[width=\columnwidth]{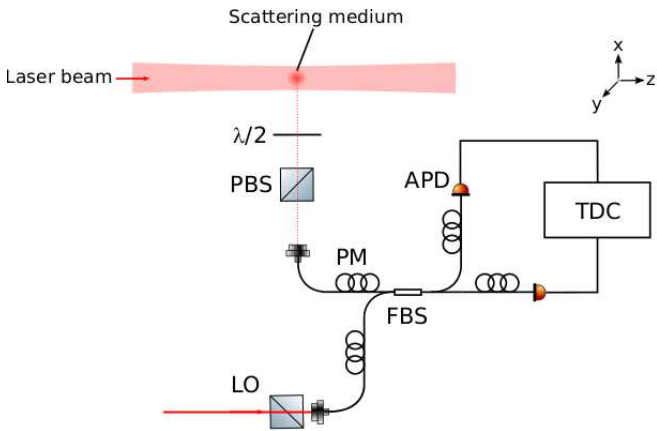}
	\caption{Experimental setup used to check the validity of the Siegert relation. The chaotic light corresponds to the light scattered by a medium illuminated by a laser beam. This scattered light is collected by a polarization maintaining (PM) single-mode fiber. Its polarization is selected by a $\lambda/2$ plate and a polarizing beam splitter (PBS). The light is finally split with a fibered beam splitter (FBS) and its two outputs illuminate two avalanche photodiodes (APDs). Each photon arrival is time-tagged by a time-to-digital converter (TDC) and the correlations are computed by a computer. Finally, a local oscillator (LO), derived form the same laser that illuminates the scattering medium, is injected in the second input of the FBS.}
	\label{fig:Setup}
\end{figure}

To test the validity of the Siegert relation, we have implemented a setup which allows measuring $g^{(1)}(\tau)$ and $g^{(2)}(\tau)$ at the same time. This configuration is particularly suited to ensure that the two measurements are done exactly under the same conditions, in order to overcome the unavoidable fluctuations from one experiment to the other. The setup is depicted in Fig.\,\ref{fig:Setup}.

\subsubsection{Chaotic light}

The Siegert relation has been derived in the previous section for photons emitted by a large number of independent emitters. Experimentally, this can be achieved, for example, by scattering laser light on a medium. This medium can be either a rotating diffuser, such as a glass diffuser or a simple piece of white paper, some milk, some scatterers immersed in a fluid, or a gas. The common feature for all these media is the time evolution of the position of the scatterers, either due to mechanical motion or temperature and Brownian motion.

In our experiment, the scattering sample corresponds to a cold atomic cloud, produced by loading a magneto optical trap (MOT) from a vapor of $^{85}$Rb. More details can be found in Refs.\,\cite{Kashanian_2016,Eloy_2018}. The cloud is made of typically a few $10^7$ atoms with an rms radius of a few hundreds of $\mu$m and a temperature ranging from 100\,$\mu$K to 10 mK. The light that illuminates the cloud comes from a distributed-feedback laser amplified by a tapered amplifier, whose frequency is locked close to the $F = 3 \rightarrow F' = 4$ hyperfine transition of the $^{85}$Rb $D_2$ line.

\subsubsection{Measurement of $g^{(2)}(\tau)$}

The general Siegert relation is given by Eq.\,(\ref{eq:Siegert_beta}). The signal is maximum with $g^{(2)}(\tau = 0) = 2$ for $\beta = 1$, which means that only one spatial and polarization mode is selected. The spatial selection is achieved by injecting the scattered light in a polarization maintaining (PM) single-mode fiber. The polarization is selected just before the fiber with a $\lambda/2$ plate and a polarizing beam splitter (PBS) and is set to be parallel to the axis of the PM fiber. The angle between the laser probe beam and the axis of the fiber was adjusted to 90$^\circ$.

Intensity correlations are then measured by sending the collected light to a detector. This can be done either in the analog regime, by recording the intensity as a function of time $I(t)$ on a photodiode, or in the single-photon counting regime as on our experiment by time-tagging the arrival of each photon with a photon-counter device. In the latter, the histogram of the time intervals between photon arrivals is computed to get the second-order correlation function $g^{(2)}(\tau)$.

Our photon counting device is an avalanche photodiode (APD) from Excelitas Technologies (SPCM-AQRH) with a fibered input. Different important parameters must be taken into account. The first one is the jitter, i.e., the temporal resolution of the detector. In order to be able to neglect this effect, one should measure signals with a frequency much lower than the inverse of its temporal resolution. Our APD has a jitter of a few hundreds of ps corresponding to a cutoff frequency of typically 1\,GHz. The second important parameter is the dead time $t_\mathrm{d}$ of the detector, of the order of 20\,ns for our device; it corresponds to the time during which, after each event, the detector is not able to record any new event. This means that no intensity correlations can be recorded for $\tau$ smaller than the dead time, and thus very close to $\tau = 0$. The last parameter is the afterpulsing, a parameter which gives the probability that the counting device produces a fake count at the end of the dead time and thus after the first detection of an event. This parameter, of the order of 1\,$\%$ on our setup, results in a technical correlation around $\tau = t_\mathrm{d}$ that adds to the signal we want to measure. To overcome these detrimental effects, one can use two APDs, as depicted in Fig.\,\ref{fig:Setup}, and measure the correlation between the two outputs. To do so, we connect the first PM fiber, which collects the scattered light, to a 50:50 fibered beam splitter (FBS). The two outputs of the FBS are used to illuminate each APD.

Finally, the arrival time of each photon recorded on the APDs is processed by a time-to-digital converter (TDC, ID800 from IDQuantique), with a time resolution of 162\,ps. The data are transferred to a computer and processed by a Matlab code to calculate the histogram of the coincidences between the two channels, with a time bin equal to the TDC time resolution.

\subsubsection{Measurement of $g^{(1)}(\tau)$}

The measurement of $g^{(1)}(\tau)$ is done with a standard optical heterodyne technique: we superimpose the light under investigation and a laser beam, denoted as local oscillator (LO) in Fig.\,\ref{fig:Setup}, and we record the beat note. In this kind of configuration, we get (see Eq.~(\ref{eq:g2_BN1}) in the following section):
\begin{equation}
   g^{(1)}(\tau) = g_\mathrm{sc}{}^{(1)}(\tau)\times g_\mathrm{LO}{}^{(1)}(\tau).
\end{equation}
Thus, the first-order correlation function is given by the product of the electric field correlation of the scattered light and the one of the LO. In Fourier space, this corresponds to the convolution of the two optical spectra. If the coherence time of the scattered light is much smaller than the coherence time of the LO, or in other words, if the laser linewidth is much smaller than the one of the scattered light, one gets: $g^{(1)}(\tau) \simeq g_\mathrm{sc}{}^{(1)}(\tau)$.

In our setup, the LO is derived from the same laser used to illuminate the scattering medium but with an extra detuning $\omega_\mathrm{BN}$ added with an acousto-optic modulator. It is injected in the second input of the FBS, with a polarization parallel to the collected scattered light. The intensity of the LO is set to maximum, just before the saturation of the APDs.

\subsection{Intensity correlation}\label{subsec:setup_correlations}

What we experimentally measure are the intensity correlations between the two output ports of the FBS and thus the intensity correlation of the beat note between the scattered light and the LO:
\begin{equation}
    g_\mathrm{BN}{}^{(2)}(\tau) = \frac{\langle I_\mathrm{BN}(t)I_\mathrm{BN}(t + \tau) \rangle}{\langle I_\mathrm{BN}(t) \rangle^2}.
\end{equation}
In the following, the electric field of the scattered light is denoted as:
\begin{equation}
    E_\mathrm{sc}(t) = E_{\mathrm{sc},0}(t)e^{{i[\omega_\mathrm{sc}t + \phi_\mathrm{sc}(t)]}} ,
\end{equation}
with $\omega_\mathrm{sc}$ the light frequency, $\phi_\mathrm{sc}(t)$ its phase noise and $I_ \mathrm{sc} = |E_{\mathrm{sc},0}(t)|^2$ its intensity noise. Similarly, we will write the LO electric field as follows:
\begin{eqnarray}
    E_\mathrm{LO}(t) e^{i\omega_\mathrm{BN}t} &=& E_{\mathrm{LO},0}(t)e^{{i[\omega_\mathrm{LO}t + \phi_\mathrm{LO}(t)]}},\nonumber\\
    &=& E_{\mathrm{LO},0}(t)e^{{i[\omega_\mathrm{sc}t + \phi_\mathrm{LO}(t)]}}e^{i\omega_\mathrm{BN}t},
\end{eqnarray}
with $E_{\mathrm{LO},0}$ the LO amplitude, $\omega_\mathrm{LO}=\omega_\mathrm{sc}+\omega_\mathrm{BN}$ its frequency and $\phi_\mathrm{LO}$ its phase.
Finally, the electric field of the beat note and its intensity are given by:
\begin{eqnarray}
    E_\mathrm{BN}(t) &=& E_\mathrm{LO}(t) + E_\mathrm{sc}(t),\\
    I_\mathrm{BN}(t) &=& I_\mathrm{LO}(t) + I_\mathrm{sc}(t) + [E_\mathrm{LO}(t) E_\mathrm{sc}^{\star}(t)e^{i\omega_\mathrm{BN}t} +c.c.], \nonumber
\end{eqnarray}
with $c.c.$ the complex conjugate.
The intensity correlation then writes:
\begin{widetext}
\begin{eqnarray}
&&\langle I_\mathrm{BN}(t)I_\mathrm{BN}(t + \tau) \rangle \nonumber\\
&=& \langle I_\mathrm{LO}(t) I_\mathrm{LO}(t+\tau) \rangle + \langle I_\mathrm{sc}(t) I_\mathrm{sc}(t+\tau) \rangle + 2\langle I_\mathrm{LO} \rangle \langle I_\mathrm{sc} \rangle
+ \langle E_\mathrm{LO}(t)E_\mathrm{LO}^\star(t+\tau)E_\mathrm{sc}(t)^\star E_\mathrm{sc}(t+\tau) e^{-i (\omega_\mathrm{BN}\tau + \pi)} + c.c. \rangle, \nonumber\\
\label{eq:II_1}\\
&=& \langle I_\mathrm{LO}(t) I_\mathrm{LO}(t+\tau) \rangle + \langle I_\mathrm{sc}(t) I_\mathrm{sc}(t+\tau) \rangle + 2\langle I_\mathrm{LO} \rangle \langle I_\mathrm{sc} \rangle + (\langle E_\mathrm{LO}(t)E_\mathrm{LO}^\star(t+\tau)\rangle \langle E_\mathrm{sc}(t)^\star E_\mathrm{sc}(t+\tau)\rangle e^{-i (\omega_\mathrm{BN}\tau + \pi)} + c.c.), \nonumber \\
\label{eq:II_2}\\
&=&\langle I_\mathrm{LO} \rangle^2 g_\mathrm{LO}{}^{(2)}(\tau)  + \langle I_\mathrm{sc} \rangle^2 g_\mathrm{sc}{}^{(2)}(\tau) + 2\langle I_\mathrm{LO} \rangle \langle I_\mathrm{sc} \rangle +  2\langle I_\mathrm{LO} \rangle \langle I_\mathrm{sc} \rangle g_\mathrm{LO}{}^{(1)}(\tau) g_\mathrm{sc}{}^{(1)}(\tau) \cos{(\omega_\mathrm{BN} \tau + \pi)}. \nonumber\\
\label{eq:II_3}
\end{eqnarray}
\end{widetext}
The first two terms of Eq.\,(\ref{eq:II_1}) correspond to the intensity correlations of the LO and of the scattered light, while the last terms are linked to electric field correlations. The terms proportional to $e^{\pm 2i \omega_\mathrm{BN}t}$ average to zero and are thus not written. The $\pi$ phase comes from the phase difference that one gets between the two outputs of the beam splitter. We have assumed that the LO field and the scattered electric field are independent quantities [Eq.\,(\ref{eq:II_2})], and that the electric field correlations are real and positive [Eq.\,(\ref{eq:II_3})]. The normalized intensity correlation function is finally given by:
\begin{widetext}
\begin{eqnarray} \label{eq:g2_BN1}
g_\mathrm{BN}{}^{(2)}(\tau) &=& 1 + 2\frac{\langle I_\mathrm{sc} \rangle \langle I_\mathrm{LO} \rangle}{\left(\langle I_\mathrm{sc} \rangle + \langle I_\mathrm{LO} \rangle \right)^2} g_\mathrm{LO}{}^{(1)}(\tau) g_\mathrm{sc}{}^{(1)}(\tau) \cos(\omega_\mathrm{BN} \tau + \pi) \nonumber \\
&+& \frac{\langle I_\mathrm{sc} \rangle^2}{\left(\langle I_\mathrm{sc} \rangle + \langle I_\mathrm{LO} \rangle \right)^2}\left(g_\mathrm{sc}{}^{(2)}(\tau)-1 \right) + \frac{\langle I_\mathrm{LO} \rangle^2}{\left(\langle I_\mathrm{sc} \rangle + \langle I_\mathrm{LO} \rangle \right)^2}\left(g_\mathrm{LO}{}^{(2)}(\tau)-1 \right).
\end{eqnarray}

This equation can be further simplified if the LO comes from a laser: $g_\mathrm{LO}{}^{(2)}(\tau) = 1$ whatever the interval $\tau$. Finally, assuming that the temporal coherence of the LO is much larger than the one of the scattered light, $g_\mathrm{sc}{}^{(1)}(\tau)$ decays much faster than $g_\mathrm{LO}{}^{(1)}(\tau)$, and one obtains\,\cite{Hong_2006}:
\begin{equation}
    g_\mathrm{BN}{}^{(2)}(\tau) \simeq 1 + 2\frac{\langle I_\mathrm{sc} \rangle \langle I_\mathrm{LO} \rangle}{\left(\langle I_\mathrm{sc} \rangle + \langle I_\mathrm{LO} \rangle \right)^2}g_\mathrm{sc}{}^{(1)}(\tau) \cos(\omega_\mathrm{BN} \tau +\pi) + \frac{\langle I_\mathrm{sc} \rangle^2}{\left(\langle I_\mathrm{sc} \rangle + \langle I_\mathrm{LO} \rangle \right)^2}\left(g_\mathrm{sc}{}^{(2)}(\tau)-1 \right). \label{eq:g2_BN}
\end{equation}
\end{widetext}
All these calculations can be done in the case where the scattering medium is illuminated by a laser derived from the same laser as the LO. One finds that the intrinsic phase noise of the laser cancels, and one gets again Eq.\,(\ref{eq:g2_BN}).

From this last formula, it is now clear that this setup allows measuring the two quantities $g_\mathrm{sc}{}^{(1)}(\tau)$ and $g_\mathrm{sc}{}^{(2)}(\tau)$ at the same time. Indeed, one can note that they are not centered at the same frequency. While $g_\mathrm{sc}{}^{(2)}(\tau)$ is centered around the DC value, $g_\mathrm{sc}{}^{(1)}(\tau)$ is centered around $\omega_\mathrm{BN}$. Taking the Fourier transform of the total signal $g_\mathrm{BN}{}^{(2)}$ allows us to separate both quantities, as long as the overlap between them is negligible. We can finally directly check the Siegert relation in the Fourier space, after shifting $\tilde{g}_\mathrm{sc}{}^{(1)}(\omega)$ back to zero frequency.

\subsection{Typical signals}\label{subsec:typical_signals}

\begin{figure}[ht]
	\centering
	\includegraphics[width=8cm]{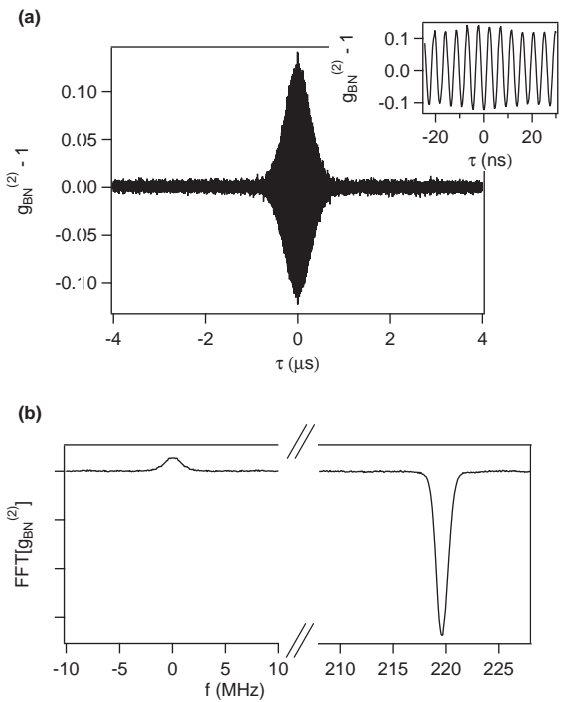}
	\caption{(a) Typical temporal intensity correlation of the beat note. Inset: Zoom around $\tau = 0$ showing the beat note between the scattered light and the LO. (b) Fourier transform of the temporal intensity correlation. The curve centered around $f=0$ corresponds to the Fourier transform of $g_\mathrm{sc}{}^{(2)}$, while the curve centered around $f = f_\mathrm{BN} \simeq 220$\,MHz corresponds to the Fourier transform of $g_\mathrm{sc}{}^{(1)}$.}
	\label{fig:g2_BN}
\end{figure}

A typical normalized intensity correlation function obtained in our experiment is plotted in Fig.\,\ref{fig:g2_BN} (a). For this specific experiment, we had $\langle I_\mathrm{LO} \rangle \simeq 8 \langle I_\mathrm{sc} \rangle$. If we zoom around $\tau = 0$, as shown in the inset of Fig.\,\ref{fig:g2_BN}, the beat note between the scattered light and the LO is clearly visible, with a frequency beat note of $\omega_\mathrm{BN}/2 \pi = f_\mathrm{BN} \simeq 220$\,MHz. The decay of the envelope gives an estimate on the coherence time of the scattered light, which is of the order of $1\,\mu$s in this example. It is however difficult to get more quantitative information since the contributions of the electric field and intensity correlations are here mixed.

To be able to extract separately the contributions of $g_\mathrm{sc}{}^{(1)}(\tau)$ and $g_\mathrm{sc}{}^{(2)}(\tau)$, we take the Fourier transform of the temporal intensity correlation of the beat note. As shown in Fig.\,\ref{fig:g2_BN} (b), these two contributions are centered around different frequencies: around 0 for the Fourier transform of the intensity correlation $\tilde g_\mathrm{sc}{}^{(2)}$ and around $f_\mathrm{BN}$ for the Fourier transform of the electric field correlation $\tilde g_\mathrm{sc}{}^{(1)}$. The difference in height comes from the difference between the number of photons detected on the detector coming from the scattered light and the LO. While $\tilde g_\mathrm{sc}{}^{(2)}$ is positive, $\tilde g_\mathrm{sc}{}^{(1)}$ is negative due to the $\pi$ phase in the second term of Eq.\,(\ref{eq:g2_BN}).

As said in Sec.\,\ref{subsec:Siegert_relation}, the Siegert relation can be checked either in the time domain or in the frequency domain. Since it is much easier to extract separately the two quantities of interest, $g_\mathrm{sc}^{(1)}$ and $g_\mathrm{sc}^{(2)}$, in Fourier space, we hereafter work in the frequency domain. To do so, we first need to check that both contributions can be well isolated. The choice of $\omega_\mathrm{BN}$ is here crucial: it should be smaller than the frequency bandwidth of the detector, but large enough to avoid any overlap between $\tilde g_\mathrm{sc}{}^{(1)}$ and $\tilde g_\mathrm{sc}{}^{(2)}$. We then isolate the two contributions $\tilde g_\mathrm{sc}{}^{(1)}$ and $\tilde g_\mathrm{sc}{}^{(2)}$. Afterwards, the term corresponding to the electric field correlation $\tilde g_\mathrm{sc}{}^{(1)}$ is frequency-centered around zero and its self-convolution is calculated. We finally superimpose the two quantities on the same graph, after normalizing their height.

\begin{figure}[ht]
	\centering
	\includegraphics[width=\columnwidth]{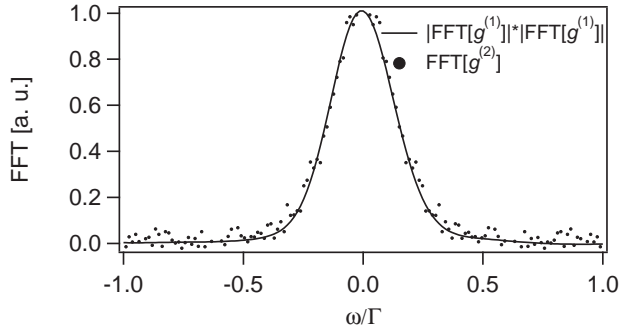}
	\caption{Siegert relation in the low saturation limit ($s(\Delta)\ll 1$) and single scattering regime ($b(\Delta)<1$). See Sec.\,\ref{subsec:Low_s_Single_scattering} for more details on the experimental parameters. Points: Fourier transform of the intensity correlation $\tilde g_\mathrm{sc}{}^{(2)}$; plain curve: self-convolution of the Fourier transform of the electric field correlation $|\tilde g^{(1)}(\omega)| * |\tilde g^{(1)}(\omega)|$.}
	\label{fig:single_scattering}
\end{figure}

This comparison is presented in Fig.\,\ref{fig:single_scattering}, for the data in Fig.\,\ref{fig:g2_BN}. The points correspond to the Fourier transform of the intensity correlation $\tilde g_\mathrm{sc}{}^{(2)}$, while the plain curve corresponds to the autoconvolution of $\tilde g_\mathrm{sc}{}^{(1)}$. We observe an excellent overlap between the two curves, validating the Siegert relation for this specific experiment. This experiment has been repeated in different kinds of configuration, as detailed in the next section.

\section{Tests in different configurations}\label{sec:Tests}

\subsection{Low-saturation limit, single-scattering regime}
\label{subsec:Low_s_Single_scattering}

The light under investigation is the light scattered by a cloud of Rubidium cold atoms shone by a laser probe beam. We have first tested the Siegert relation with a low saturation parameter. This ensures that photons are mainly scattered elastically. In this configuration, the frequency of the scattered light remains the same as the incident one, apart from a Doppler shift due to the velocity of the could center of mass and the temperature\,\cite{Eloy_2018}. The intensity $I$ of the probe is adjusted in order to have a saturation parameter $s \simeq 0.1$, where
\begin{equation}
s = \frac{I/I_\mathrm{sat}}{1+4\Delta^2/\Gamma^2},
\end{equation}
with $I_\mathrm{sat} \simeq 3.7$\,mW/cm$^2$, $\Gamma/2\pi = 6.07$\,MHz the linewidth of the $D_2$ hyperfine transition and $\Delta$ the laser detuning compared to the $F = 3 \rightarrow F' = 4$ transition.

Another important parameter that can be controlled in the experiment is the optical thickness of the cloud
\begin{equation}
    b(\Delta) = \frac{b_0}{1+4\Delta^2/\Gamma^2},
\end{equation}
with $b_0$ the optical thickness at resonance. By varying the detuning of the probe, one can explore either the single-scattering regime, where $b(\Delta)<1$ and for which an incident photon is scattered at most once before escaping the cloud, or the multiple-scattering regime, when $b(\Delta)>1$ and for which an incident photon can be multiply scattered before escaping the cloud. The impact of this transition on the temporal intensity correlations and in particular on the coherence time of the light has been discussed in detail in Ref.\,\cite{Eloy_2018}.

For the first experiment, the on-resonance optical thickness has been adjusted to $b_0 \simeq 20$. The detuning was set to $\Delta = 4\Gamma$ in order to have $b(\Delta) \simeq 0.3$ and thus to be in the single scattering regime. We also applied two counter-propagating probe beams instead of one, in order to not push the atoms. The results are plotted in Fig.\,\ref{fig:single_scattering}, showing a very good overlap between the contributions of the $g_\mathrm{sc}{}^{(2)}$ and $g_\mathrm{sc}{}^{(1)}$ correlation functions. The validity of the Siegert relation is thus confirmed under these conditions. The full width at half maximum (FHWM) of $\tilde g^{(2)}(\omega)$ is of the order of $2\pi\times 2$\,MHz, corresponding to a coherence time of 190\,ns and to a temperature of about 1\,mK\,\cite{Eloy_2018}. This temperature is much higher than the one measured on the cloud before the application of the probe beam (100\,$\mu$K). This increase is actually due to the photons exchange during the application of the probe beam, which heats up the atomic sample.

\subsection{Low-saturation limit, multiple-scattering regime}
\label{subsec:low_s_multiple}

\begin{figure}[ht]
	\centering
	\includegraphics[width=\columnwidth]{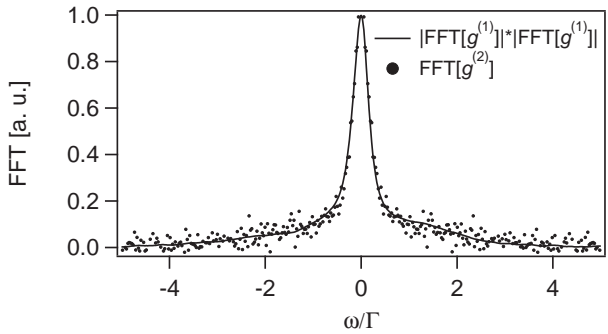}
	\caption{Siegert relation in the low saturation limit, $s(\Delta)\ll 1$, and multiple-scattering regime, $b(\Delta)>1$. See Sec.\,\ref{subsec:low_s_multiple} for more details. Points: Fourier transform of the intensity correlation $\tilde g_\mathrm{sc}{}^{(2)}$; plain curve: self-convolution of the Fourier transform of the electric field correlation $|\tilde g^{(1)}(\omega)| * |\tilde g^{(1)}(\omega)|$.}
	\label{fig:multiple_scattering}
\end{figure}

We then turned to the multiple scattering regime, but still with a low saturation parameter. The detuning is smaller, set to $\Delta = 1.5\Gamma$. The new optical thickness is $b(\Delta) = 2$. The test of the Siegert relation is plotted in Fig.\,\ref{fig:multiple_scattering} and we again see a very good overlap between the two curves. The main difference with Fig.\,\ref{fig:single_scattering} comes from the coherence time, which is reduced compared to the single scattering regime, and thus to an increase of the frequency linewidth, with a FWHM on $\tilde g^{(2)}(\omega)$ of $2\pi \times 2.6$\,MHz. This is a direct signature of the frequency redistribution induced by multiple scattering.

\subsection{Strong saturation limit}
\label{subsec:large_s}

The last set of measurements has been done in the single scattering regime with a laser probe at resonance and with a strong intensity. In this regime, the light is mainly scattered inelastically and its spectrum corresponds to the well-known Mollow triplet\,\cite{Mollow_1969}: it presents two sidebands distant from a carrier by the Rabi frequency. The experimental details, the time sequence and the specific study of this kind of spectra can be found in Ref.\,\cite{Ortiz_2019}. The saturation parameter is of the order of $s \simeq 130$, corresponding to a Rabi frequency at the centre of the probe beam of $\Omega \simeq 8\Gamma$.

\begin{figure}[ht]
	\centering
	\includegraphics[width=\columnwidth]{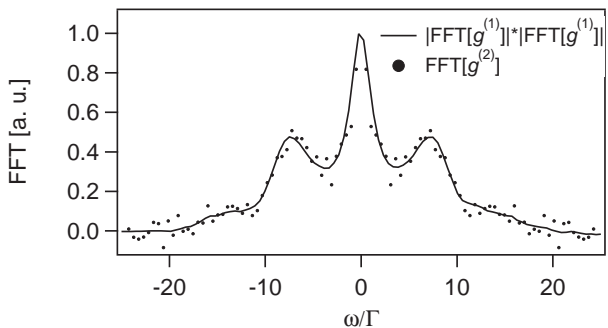}
	\caption{Siegert relation in the strong saturation limit ($s(\Delta)>1$). The peaks of the Mollow triplet are clearly visible. Points: Fourier transform of the intensity correlation $\tilde g_\mathrm{sc}{}^{(2)}$; plain curve: self-convolution of the Fourier transform of the electric field correlation $|\tilde g^{(1)}(\omega)| * |\tilde g^{(1)}(\omega)|$.}
	\label{fig:Mollow}
\end{figure}

The results presented in Fig.\,\ref{fig:Mollow} show that the Siegert relation is also valid in this situation. The Mollow triplet is clearly visible. Since the Fourier transform of $g_\mathrm{sc}{}^{(2)}$ corresponds to the self-convolution of the light spectrum made of three peaks, two more peaks appear on each side of the curve at $2\Omega$ from the carrier. These are visible on $|\tilde g^{(1)}(\omega)| * |\tilde g^{(1)}(\omega)|$, and should also appear on $\tilde g^{(2)}(\omega)$. However, they are hidden by the noise. This illustrates the fact that even if we get the same signals due to the validity of the Siegert relation, the signal to noise ratio is very different for each type of measurement. While this ratio can be improved for $g^{(1)}(\tau)$ by simply increasing the intensity of the LO, in the case of $g^{(2)}(\tau)$ it is limited by the amount of scattered light that one can collect. This effect can also be observed for the other configurations discussed in this paper (see Figs.~\ref{fig:single_scattering} and~\ref{fig:multiple_scattering}).

\section{Conclusion}
\label{sec:Conclusion}
In this paper, we have presented an experimental setup to test the validity of the Siegert relation. This can be easily adapted for a student laboratory experiment. In particular, instead of studying the light scattered by a cold atomic cloud, which requires a complicated and rather expensive setup, one can use other sources of pseudothermal light such as the light scattered by a rotating ground disk illuminated by a laser\,\cite{Martienssen_1964,Haner_1970,Kuusela_2017}. This solution is easy and low-cost to implement.

In our specific configuration, we have tested the Siegert relation in the case of a laser beam scattered by a large cloud of cold atoms, from the single to the multiple scattering regime and for small and high laser intensity. The validity of this relation illustrates the fact that the scatterers are essentially uncorrelated, which is the frame of the derivation presented in Sec.\,\ref{sec:Siegert_derivation}. This applies whether the atoms behave as classical scatterers (weak laser intensity) or as quantum emitters (high laser intensity).

This derivation relies on two main assumptions: a large number $N$ of emitters as shown by Eqs.\,(\ref{eq:g2N}) and \,(\ref{eq:Siegert_simple_bis}), and independent or uncorrelated emitters. When $N$ is small, the intensity correlation function depends on $N$. This property can actually be used to evaluate the number of scatterers~\cite{Watts1996}, or even its fluctuations~\cite{Schaefer1972}. In the extreme limit of a single quantum scatterer ($N=1$), the classical description is not even valid anymore, as illustrated by the phenomena of photon antibunching for a single-atom~\cite{Paul1982,Kimble1977} for which $g^{(2)}(0)=0$.

The hypothesis of random independent phases emitted by the different scatterers is also an important ingredient of the derivation, and it has already been shown that the presence of correlations between the scatterers yields deviations from the Siegert relation~\cite{Voigt1994}. In the context of multiple scattering, the shortest scattering paths~\cite{Page2003} (i.e., which involve particles with correlated motion) or the presence of ``static paths''\,\cite{Boas1997,Borycki2016} (i.e., which do not change over time) bring again deviations to the Siegert relation.

Finally, the validity or the violation of the Siegert relation is an important tool to get informations on the nature of a light source, as demonstrated on many different kinds of sources going from light scattered on samples of polystyrene spheres~\cite{Pine1990,Lemieux1999} or foam~\cite{Gittings2006}, pseudo-thermal source\,\cite{Kuusela_2017} or laser light~\cite{Stevens2010}. As a final remark, we remind that it addresses only the lowest optical coherences, i.e. $g^{(1)}$ and $g^{(2)}$, and that a complete characterization of light source properties requires the study of the correlation function $g^{(n)}$ at all orders~\cite{Lemieux1999}.

\begin{acknowledgments}
Part of this work was performed in the framework of the European Training Network ColOpt, which is funded by the European Union (EU) Horizon 2020 programme under the Marie Sklodowska-Curie action, grant agreement No.~721465. M.F., R.B.~and R.K.~received support from project CAPES-COFECUB (Ph879-17/CAPES 88887.130197/2017-01). R.B. benefited from grants from the S\~ao Paulo Research Foundation (FAPESP) (Grants Nrs.~18/01447-2, 18/15554-5 and 19/13143-0). This study was supported by the Coordena\c{c}\~{a}o de Aperfei\c{c}oamento de Pessoal de N\'ivel Superior - Brasil (CAPES) - Finance Code 001, the Excellence Initiative UCA-JEDI from University C\^ote d'Azur, and the French National Research Agency (ANR19-CE47-0014-01).
\end{acknowledgments}

\bibliography{Biblio,HBT_paper_biblio}




\end{document}